\documentclass[sigconf]{acmart}

\usepackage{todonotes}
\usepackage{outlines}
\usepackage{amssymb}
\usepackage{listings}

\setcopyright{acmlicensed}

\acmDOI{http://dx.doi.org/10.1145/3067695.3084216}

\acmISBN{978-1-4503-4939-0/17/07}
\acmPrice{15.00}

\acmConference[GECCO '17]{the Genetic and Evolutionary Computation Conference 2017}{July 15--19, 2017}{Berlin, Germany}
\acmYear{2017}
\copyrightyear{2017}

\begin{document}
\tolerance=10000

\title{TensorFlow Enabled Genetic Programming}

\author{Kai Staats}
\affiliation{%
  \institution{Embry-Riddle Aeronautical Univ.}
  \city{Prescott} 
  \state{Arizona}
	\country{USA}
}

\author{Edward Pantridge}
\affiliation{
	\institution{MassMutual Financial Group}
	\city{Amherst}
	\state{Massachusetts}
	\country{USA}
}

\author{Marco Cavaglia}
\affiliation{
	\institution{University of Mississippi}
	\city{University}
	\state{Mississippi}
	\country{USA}
}

\author{Iurii Milovanov}
\affiliation{
	\institution{SoftServe, Inc.}
	\city{Austin}
	\state{Texas}
	\country{USA}
}

\author{Arun Aniyan}
\affiliation{
	\institution{SKA South Africa / Rhodes University}
	\city{Cape Town}
	\state{WC}
	\country{South Africa}
}

\begin{abstract}
Genetic Programming, a kind of evolutionary computation and machine learning algorithm, is shown to benefit significantly from the application of vectorized data and the TensorFlow numerical computation library on both CPU and GPU architectures. The open source, Python \textit{Karoo GP} is employed for a series of 190 tests across 6 platforms, with real-world datasets ranging from 18 to 5.5M data points. This body of tests demonstrates that datasets measured in tens and hundreds of data points see 2-15x improvement when moving from the scalar/SymPy configuration to the vector/TensorFlow configuration, with a single core performing on par or better than multiple CPU cores and GPUs. A dataset composed of 90,000 data points demonstrates a single vector/TensorFlow CPU core performing 875x better than 40 scalar/Sympy CPU cores. And a dataset containing 5.5M data points sees GPU configurations out-performing CPU configurations on average by 1.3x.
\end{abstract}

\ccsdesc[500]{Software and its engineering~Genetic programming}
\ccsdesc[500]{Software and its engineering~Massively parallel systems}

\keywords{genetic programming, evolutionary computation, machine learning, multicore, gpu, tensorflow, vectorized, parallel}

\maketitle

\section{Parallel Computing}
\subsection{A brief introduction}
Optimizing code for parallel execution has since the advent of computer modeling and data processing been integral to high performance computing. Parallelism comes in many forms, and functions at various scales, from multiple Central Processing Unit (CPU) cores on a single microprocessor chip to multiple microprocessors on a card or motherboard; from multiple motherboards tightly coupled by an internode communication fabric to widely distributed cloud computing.

With the introduction of the General Purpose Graphics Processing Unit (GPGPU or GPU), massively parallel execution in a small footprint is made possible. Where multicore CPU configurations commonly incorporate 2, 4, or 8 cores, GPUs tightly integrate hundreds, even thousands of processing cores per card. Initially designed to drive realistic, real-time gaming engines, the GPU has enabled the growing machine learning community to build computationally powerful systems such as Deep Learning Algorithms (DLA) \cite{NIPS2012_4824}.

However, the common claim that GPUs perform 10x to 1000x better than CPUs is proved to be application specific. Lee et al. \cite{lee2010debunking} demonstrate that while CPU architectures are limited by the number of cores which fit onto a single die, and the number of chips per motherboard, CPUs do retain some advantages over GPUs by means of a higher clock rate, larger cache, and inherent design to work with a far greater variety of applications than GPUs. Furthermore, the increased core count of GPUs results in increased overhead.

Performance metrics of both CPUs and GPUs are often based upon the data residing entirely on-card. Gregg and Hazelwood \cite{gregg2011data} demonstrate that to compare CPU to GPU performance without first specifying the load and return time of the data is to ignore the fact that movement of data onto a GPU card can be 2-50x greater than the processing time alone.

While it is not the intent of this paper to provide a detailed analysis of the hardware level function of microprocessors, this basic understanding works to explain the otherwise surprising outcome of some of the tests described herein.


\subsection{Genetic Programming in Parallel}
Genetic Programming (GP) has since the mid 1990s seen effort to improve computational performance through the application of parallel processing. Dozens of publications are found at the GP Bibliography\footnote{liinwww.ira.uka.de/bibliography} which discuss parallelization of GP on both CPU and GPU architectures. In most cases, the data and/or populations are divided for simultaneously processing. Darren M. Chitty \cite{chitty2012fast} provides a comprehensive summary of more than 30 efforts conducted on transputers, FPGAs, XBox 360s, and GPUs. Cano et al. demonstrate an 820x increase in performance on Nvidia GPUs \cite{cano2012speeding} while Augusto and Barbosa discuss OpenCL as a means for a wide range of evolutionary algorithms to take advantage of GPU cores \cite{augusto2013accelerated}.

The diversity of these efforts make clear that Genetic Programming is inherently able to scale to multicore and manycore architectures and achieve a noteworthy increase in performance. While code written in C++ and NVidia's CUDA library can enable an optimal level of performance, libraries such as TensorFlow, Theano, Caffe, and Torch provide developers the ability to take advantage of single computer, multicore CPU and GPU architectures without having to master a more advanced understanding and associated routines \cite{tensorflow2015-whitepaper}.

\section{Karoo GP}
\subsection{Introduction}
\textit{Karoo GP}\footnote{kstaats.github.io/karoo\_gp/} is a Genetic Programming suite written in the computer language Python. Karoo GP is highly scalable, with vectorized data, multicore CPU and GPU support by means of the TensorFlow library. Karoo GP was developed by Kai Staats in the course of his MSc research to analyze data produced at the Square Kilometre Array---South Africa (SKA-SA).

In its first deployment, Karoo GP was engaged in the mitigation of radio frequency interference in radio astronomy data at the SKA-SA. Karoo GP is now employed in data analysis at the Laser Interferometer Gravitational-wave Observatory (LIGO) for the glitch classification and subsequent effort to better understanding associated, real-world mechanical couplings. Karoo GP is also employed at LIGO in an early-stage effort to classify supernovae injections from background noise, and in neutrino detector data analysis at the Ohio State University Center for Cosmology and Astroparticle Physics (CCAPP).

Karoo GP includes an intuitive desktop and scriptable, command-line user interface. All configuration parameters and populations are automatically archived. The included User Guide offers system requirements, a crash-course in Genetic Programming, and use of Karoo GP for both the novice and advanced user.

\subsection{Key Features}
\begin{itemize}
\item Written in Object Oriented Python with a hierarchical naming scheme for all methods.
\item Multicore CPU and GPU support through the TensorFlow library.
\item Desktop interface provides a text-based menu, five display modes, and runtime reconfiguration of parameters.
\item Server interface enables scriptable runs via command-line arguments.
\item Anticipates standard Comma Separated Value datasets.
\item Automatically archives the population and configuration parameters of each generation.
\item Supports customized seed populations.
\item Simple framework for preparing custom fitness functions and evaluation routines.
\end{itemize}

\subsection{The Code}
Karoo GP is built upon the classic tree-based approach to GP, as described in the first three chapters of the "Field Guide to Genetic Programming" by Poli, Langdon, McPhee, and Koza \cite{poli2008field}. At a high level, the code is described by seven families of Python methods, as provided in Table 1.

\begin{table}[!h]
\begin{tabbing}
\hspace*{6mm}\textbf{Method Family} \hspace*{6mm} \= \textbf{Description} \\
\hspace*{6mm}fx\_karoo\_ \> Methods to Run Karoo GP \\
\hspace*{6mm}fx\_gen\_ \> Methods to Generate a Tree \\
\hspace*{6mm}fx\_eval\_ \> Methods to Evaluate a Tree \\
\hspace*{6mm}fx\_fitness\_ \> Methods to Train and Test a Tree \\
\hspace*{6mm}fx\_evolve\_ \> Methods to Evolve a Population \\
\hspace*{6mm}fx\_display\_ \> Methods to Display a Tree \\
\hspace*{6mm}fx\_archive\_ \> Methods to Archive \\
\end{tabbing}
\caption{The Python method families in Karoo GP}
\end{table}

Of these families, two contain methods which lend themselves to parallelism: \textit{Evaluate} and \textit{Train and Test}. The other methods are engaged primarily in tournament selection and genetic operations (reproduction, mutation, cross-over). In comparison to the evaluation of an evolved expression against the data, this bookkeeping is computationally inexpensive.

The first version of Karoo GP (November 2015) included the multicore library pprocess\footnote{pypi.python.org/pypi/pprocess}, an alternative to the standard Python \textit{multiprocessing} library. Execution of the multivariate expression generated by each tree was conducted by SymPy\footnote{www.sympy.org}, a Python library for symbolic mathematics. While flexible and simple to implement, SymPy is written entirely in Python and does not provide extensive support for vectorized data.

With the release of Karoo GP v1.0, pprocess and SymPy were replaced with TensorFlow, an open source software library for numerical computation which uses data flow graphs (Section 2.6) \cite{tensorflow2015-whitepaper}. Originally developed by researchers and engineers at the Google Brain Team, C++ based TensorFlow offers powerful, flexible support of both CPUs and GPUs on desktop, laptop, and server platforms. The performance improved 875x with the KAT-7 dataset, a reduction of average wall time from 48 hours to just over three minutes (Figure 3).

Surprisingly, this increase in performance is achieved with a \textit{reduction} of engaged cores, from 40 Intel Xeon CPU cores to a single Intel core i7. A further test with a substantially larger dataset demonstrates that GPU cards provide noteworthy performance improvements when the total number of data points (instances x features) number in the millions (see Figures 4 and 5), as one might anticipate given the massive datasets frequently processed by Deep Learning models on GPU architectures.

With limited recoding, TensorFlow provides Karoo GP the capacity to engage massive datasets on single, multicore, and GPU architectures, thereby demonstrating an expanded potential for Python-based GP across a greater Machine Learning field.

\subsection{Workflow}
In the following, we describe the evaluation of a tree as an example for how TensorFlow is employed, as found in \textit{Methods to Evaluate at Tree}. The code found in \textit{Methods to Train and Test} mirrors this core functionality and will not be described in any additional detail in this document.

Following the classic tree-based GP model, the Karoo GP workflow may be described as follows \cite{poli2008field}:

\begin{enumerate}
\item Build the initial population.
\item Evaluate each tree for its fitness score.
\item Select trees to be passed to the subsequent generation.
\item Apply genetic operators as a new generation is constructed.
\item Repeat until the code termination criteria is met.
\end{enumerate}

Step 2 is where Karoo GP is optimized for parallel processing, each tree evaluated against the entire body of data, as described in the following.

\subsection{GP Tree Evaluation with TensorFlow}
In the Karoo GP method fx\_eval\_poly, a recursive function extracts from each tree a string which contains the multivariate expression. The computational burden, the one which benefits most from parallelization, is the evaluation of the multivariate expression derived from each GP tree against the entire training dataset. This is conducted by means of TensorFlow which returns both \textit{results} and a \textit{fitness score}.

The fx\_fitness\_expr\_parse method (called from fx\_fitness\_eval) is engaged to both parse the multivariate expression and convert it into a TensorFlow operations \textit{graph}. This graph is then processed in an isolated TensorFlow session to compute the results and corresponding fitness scores.

The underlying TensorFlow routine consists of a series of operations and transformations that ultimately aim to parallelize and distribute the computations across the employed hardware, CPU and/or GPU-based.

\subsection{TensorFlow Vectorized Processing}
TensorFlow processing begins with the allocation of a constant TensorFlow vector for each of the variables in the training dataset.

Where a typical dataset might contain:
\begin{equation}
\left[\begin{array}{cc} a1,b1,c1 \\ a2,b2,c2 \\ a3,b3,c3 \end{array}\right]
\end{equation}

The data is transformed to:
\begin{equation}
\left[\begin{array}{cc} a1,a2,a3 \\ b1,b2,b3 \\ c1,c2,c3 \end{array}\right]
\end{equation}

Now, the computation of $ a^2 + c / b $ can be conducted in parallel, each column a unique vector \cite{tensorflow2015-whitepaper} that is sequentially distributed and processed, per the multivariate expression for each GP tree.

Next, TensorFlow leverages the Karoo GP 'fx\_fitness\_expr\_parse' method which relies upon the built-in Python Abstract Syntax Trees (AST) library where it transforms the input multivariate expression (i) into the Python abstract syntax grammar, and (ii) into the TensorFlow computational graph. Each variable name is assigned to the corresponding constant node and each binary/unary operation is translated into the corresponding vectorized TensorFlow function node (e.g. tf.add(), tf.multiply()).

In addition, a number of sub-graphs are created to compute tree fitness value and other operational subroutines (i.e. computing labels for the Classification kernel). Karoo GP includes a separate fitness calculation sub-routine for each of the supported kernel types (Regression, Classification, Match) that compute the deviation of the result values from the given solution in a manner unique to each fitness function.

The resulting Directed Acyclic Graph consists primarily of two types of nodes:
\begin{itemize}
\item Feeding nodes: placeholders for all terminals in the dataset
\item Output nodes: values that should be retrieved from the graph
\end{itemize}

All other nodes are considered to be intermediate unless the user attempts to read their values.

\section{The Tests}
\subsection{Background}
Using beta versions of Karoo GP (Section 2.3), processing of a KAT-7 dataset at the Square Kilometre Array, South Africa, with GP configuration parameters tree depth 5, 100 trees per generation, and 30 generations required, on average, 48 hours to complete \cite{staats2016genetic}.

While the results were noteworthy, providing 90\% average Precision-Recall for the isolation of RFI (man-made noise) in radio astronomy data, the low-level computational performance became a barrier to entry when applied to additional research, such as glitch classification at LIGO (Section 3.5). Therefore, the TensorFlow library was in late 2016 introduced as a replacement for both sympy.subs maths processor and the multicore pprocess library, providing both vectorized processing and scalable parallel computation across CPU and GPU architectures.

\subsection{Configuration of the Test Matrix}
For this body of research we employ two small, classic machine learning datasets and two larger, real-world datasets (Section 3.5), each evaluated on two to six test platform configurations (Section 3.4), 10 times each, for a total of 190 test runs.

The stop parameter was set by the quantity of generations (Section 3.3) with no early terminations.

The quality (fitness) of the evolved functions were not tested.

The random functions were not set by a seed, thereby allowing exploration of the full solutions space without inadvertent, higher or lower performance than would be experienced in a real-world run.

Tests were run at times when no other users were active or when the additional activity was at a minimum on each test platform.

\subsection{Configuration of Karoo GP}
Two versions of Karoo GP are employed: v0.9.1.6 was the last stable version with sympy.subs and pprocess prior to the upgrade to TensorFlow. Version 1.0.3 introduced TensorFlow. Both are configured with the same user-defined run-time parameters, as given in Table 2.

\begin{table}[!h]
\begin{tabbing}
\hspace*{4mm}\textbf{Configuration} \hspace*{14mm} \= \textbf{Setting} \\
\hspace*{4mm}kernel \> (c)lassify or (r)egression \\
\hspace*{4mm}tree type \> (r)amped half/half \\
\hspace*{4mm}tree depth base \> 5 \\
\hspace*{4mm}tree depth max \> 5 \\
\hspace*{4mm}min nodes and leaves \> 3 \\
\hspace*{4mm}tree pop max \> 100 \\
\hspace*{4mm}tournament size \> 10 \\
\hspace*{4mm}generation max \> 30 \\
\hspace*{4mm}floating point precision \> 4 \\
\hspace*{4mm}genetic operators: \> \\
\hspace*{6mm}reproduction \> 10\% \\
\hspace*{6mm}mutation \> 20\% \\
\hspace*{6mm}crossover \> 70\% \\
\end{tabbing}
\caption{Karoo GP configuration parameters for all tests}
\end{table}

The \textit{tree depth max} configuration parameter provides a ceiling to the evolved tree depths, inhibiting bloat. The same base and max tree depth settings were maintained across all tests to provide as close a comparison as is possible.

\subsection{Hardware}
Four computer systems are employed, as follows:
\begin{itemize}
\item Apple MacBook Pro (MBP) with 4 Intel i7 CPU cores; Ubuntu via VMWare
\item Apple MacBook Pro (MBP) with NVIDIA GeForce GT 650M, 384 cores; OSX
\item SKA-SA Intel Xeon E5-2650 server, 40 CPU cores on a single motherboard; Ubuntu
\item LIGO Intel Xeon E5-2698 cluster node with NVidia Tesla P100-SXM2 GPU card, 3584 cores; Ubuntu
\end{itemize}

$ $

As applied in Table 4 and Figures 1-3, the above computer systems in combination with various software configurations provide a total of six test platforms:

\begin{itemize}
\item \textit{1-CPU\_SP}: MBP with 1 CPU core enabled via VMWare; SymPy maths library, scalar data \newline

\item \textit{1-CPU\_TF}: MBP with 1 CPU core enabled via VMWare; TensorFlow maths library, vectorized data \newline

\item \textit{40-CPU\_PP}: SKA-SA server with 40 CPU cores; SymPy maths library, \textit{pprocess} multicore library, scalar data \newline

\item \textit{40-CPU\_TF}: SKA-SA server with 40 CPU cores; TensorFlow maths library, vectorized data \newline

\item \textit{384-GPU\_TF}: MBP with 384 GPU cores; TensorFlow maths library, vectorized data \newline

\item \textit{3584-GPU\_TF}: LIGO cluster node with 3584 GPU cores; TensorFlow maths library, vectorized data
\end{itemize}

$ $

The number of CPU cores on the MacBook Pro were controlled via VMWare through which Ubuntu is run. As VMWare is a Type-1 virtual machine, the hypervisor is running at the hardware level---same as the native OSX for the MacBook Pro GPU tests.

Karoo was \textit{not} tested across a tightly coupled parallel cluster nor a distributed system such as a cloud. All tests are conducted on single system boards of various CPU and GPU architectures.

\subsection{Datasets}
We employ four datasets, as described in Table 3:

\begin{table}[!h]
\begin{tabbing}
\hspace*{4mm}\textbf{Dataset} \hspace*{14mm} \= \textbf{Dimensions}  \hspace*{6mm} \= \textbf{Data points} \\
\hspace*{4mm}Kepler \> 9 x 2 \> 18 \\
\hspace*{4mm}Iris \> 150 x 4 \> 600 \\
\hspace*{4mm}SKA-SA KAT-7 \> 10,000 x 9 \> 90,000 \\
\hspace*{4mm}LIGO Glitch \> 4000 x 1373 \> 5,492,000 \\
\end{tabbing}
\caption{The four datasets employed: two classic Machine Learning tests; two real-world datasets.}
\end{table}

\begin{enumerate}
\item Kepler 3rd Law of Planetary Motion: \\
In the early 1600s Johannes Kepler proposed three laws of planetary motion derived from data obtained from his mentor Tycho Brahe \cite{kepler1937astronomia}. The Law of Harmonies, Kepler's 3rd law of planetary motion compares the orbital period $ p $ to the average orbital radius $ r $ of one planet to one or more other planets in orbit around the Sun. The relationship established is $ p^{2} / r^{3} $. In this simple dataset, each of the nine planets (including forsaken Pluto) are described by two features, years and Astronomical Units (AU) respectively \cite{nasa2005kepler}. Kepler's 3rd Law has become a classic regression test for machine learning algorithms. \newline

\item Iris dataset: \\
In the 1920's, botanists collected measurements on the sepal length, sepal width, petal length, and petal width of 150 iris flower specimens, 50 from each of three species: Iris Setosa, Iris Versicolour, and Iris Virginica. In 1936 R.A. Fisher hand-derived a mathematical function which separated these three species \cite{fisher1936use}. As noted at the University of California's Machine Learning Repository, this is referred to as the classic Iris multivariate dataset \cite{uci_iris_data} and is a must-solve problem for developers of classification algorithms. \newline

\item RFI mitigation with the KAT-7 radio telescope array: \\
For each integration time in radio astronomy, analogous to an optical telescope's exposure of the sky, the radio telescope captures data which is processed at sequential stages, including iterative flagging of noise prior to imaging by the astronomer. Man-made noise, Radio Frequency Interference (RFI) presents a barrier to radio astronomers much as light pollution reduces the effective capacity for optical astronomy to produce clean images.

KAT-7 is an engineering prototype hosted by the Square Kilometre Array, South Africa. This data was produced by KAT-7. In its final design form, the Square Kilometre Array is projected to produce more than a petabyte of data per day. Therefore it is imperative that Machine Learning be applied to automate the mitigation of RFI in the data acquisition and processing pipeline \cite{staats2016genetic}. \newline

\item Classification of glitches at LIGO: \\
The Laser Interferometer Gravitational-wave Observatory was on September 14, 2015 successful in the first direct detection of gravitational waves and the observation of two merging black holes \cite{abbott2016observation}. As the LIGO instruments are incredibly sensitive, short-lived bursts of environmental and instrumental noise (glitches) may adversely affect searches for gravitational waves from transient astrophysical sources.

The Detector Characterization group at LIGO is exploring Machine Learning as a tool for classification of non-astrophysical noise, correlating many of the 200,000 degrees of freedom in the instrument.

This applied LIGO dataset is simulated, composed of 2000 instances of one type of glitch and 2000 instances of all others, a foundation for ensemble, binary classification. The features are extracted from typical LIGO detector characterization analysis (peak frequency, amplitude, duration, etc.) for $ n $ auxiliary sensor channels.
\end{enumerate}

\section{Results}
This series of tests clearly demonstrates that replacing scalar data with vectorized data, and a Python maths library with a C++ based maths library provides substantial improvement in a Python-based, Genetic Programming algorithm. Furthermore, the benefit of processing with massively parallel GPU cores is demonstrated when analyzing a dataset whose total number of data points (instances x features) exceeds five million.

For smaller datasets, such as Kepler's 3rd Law of Planetary Motion whose total number of data points is just 18 (9 x 2), employment of vectorized data and the TensorFlow library provides a noteworthy improvement in performance on CPU cores. As shown in Figure 1, the average wall time is reduced from 190 seconds to 126 seconds on an Intel Xeon 40 core system, and from an average 75 seconds to 40 seconds on a single Intel i7 core. The GPU cards tested were slower than the single core, yet faster than the 40 core server.

For the Iris dataset with 600 data points (150 x 4), Karoo GP sees a substantial improvement in performance, from a single CPU core with scalar data and a Python maths library requiring an average 2241 seconds to just 143 seconds with the same CPU running TensorFlow, as shown in Figure 2.

In the case of the KAT-7 dataset (10,000 x 9), Karoo GP requires an average 48 hours on a 40 CPU Intel Xeon server with scalar data and Python-based maths processor. However, with vectorized data and a C++ maths library, Karoo GP requires an average 197 seconds to complete the same task on a single Intel i7 CPU core for an increase in performance of 875x, as shown Figure 3. The massively parallel GPU cards do not demonstrate their full potential when engaging a dataset of just 90,000 data points.

With the LIGO Glitch dataset, Karoo GP analyzes 5.5M data points (4,000 x 1,373), the largest of the four datasets. The fastest, overall test platform is the LIGO GPU cluster node which employs a 3584 core GPU card running Karoo GP with TensorFlow. In Figure 4 we see a single core and quad core system perform nearly as well as the massively parallel GPU card. This comparison begins to demonstrate how larger datasets do benefit from the computational potential of a massively parallel architecture.

It is anticipated that with future tests of TensorFlow enabled Karoo GP against even larger datasets, we will see GPUs cards provide a greater increase in performance over single CPU and multicore CPU hardware configurations. \newline

\begin{table*}[!h]
\begin{tabbing}
\hspace*{26mm} \= \textbf{1-CPU\_SP} 
\hspace*{6mm} \= \textbf{1-CPU\_TF}
\hspace*{4mm} \= \textbf{40-CPU\_PP}
\hspace*{4mm} \= \textbf{40-CPU\_TF}
\hspace*{4mm} \= \textbf{384-GPU\_TF}
\hspace*{4mm} \= \textbf{3584-GPU\_TF} \\
\hspace*{0mm}  \>  \>  \>  \>  \>  \> \\
\hspace*{0mm}Kepler dataset \> 75 \> 40 \> 190 \> 126 \> 66 \> 77 \\ 
\hspace*{0mm}Iris dataset\> 2241 \> 143 \> 1106 \> 171 \> 135 \> 153 \\
\hspace*{0mm}KAT-7 dataset \> * \> 197 \> 172800 \> 179 \> 166 \> 203 \\
\hspace*{0mm}LIGO dataset \> * \> 5485 \> ** \> ** \> n/t \> 4013 \\
\end{tabbing}

\caption{A comparison of the average of 10 wall times (seconds) for the processing of each of 4 datasets. The most significant improvement in performance is with the replacement of scalar data and a Python-based maths library (SymPy) with vectorized data and a C++ maths library (TensorFlow). Per the *, the KAT-7 dataset would require roughly 160 hours to process on a single CP (and the LIGO dataset more than one year). While the comparison would yield a 3200x improvement in performance for the KAT-7 dataset, this was deemed irrelevant. Per the **, the LIGO data could be made available only to LIGO Scientific Collaboration members and was therefore not tested on all platforms.}
\end{table*}

\begin{figure*}[!h]
\centering \includegraphics[scale=0.9]{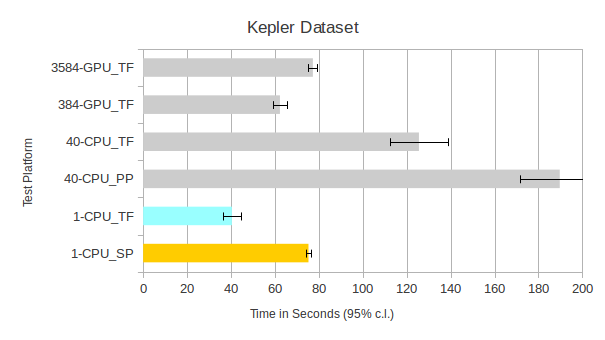}
\caption{A comparison of the average of 10 wall times (seconds) for the processing of the Kepler dataset against six test configurations (95\% cl). Of particular interest is 1-CPU\_SP vs. 1-CPU\_TF for a 2x improvement in performance from scalar data and the SymPy maths library to vector data and the TensorFlow numerical computation library. Overall, the single CPU core running TensorFlow provides the highest performance, likely due to the lack of efficiency in parallel processing such a small (9 x 2) dataset.}
\end{figure*}

\begin{figure*}[!h]
\centering \includegraphics[scale=0.9]{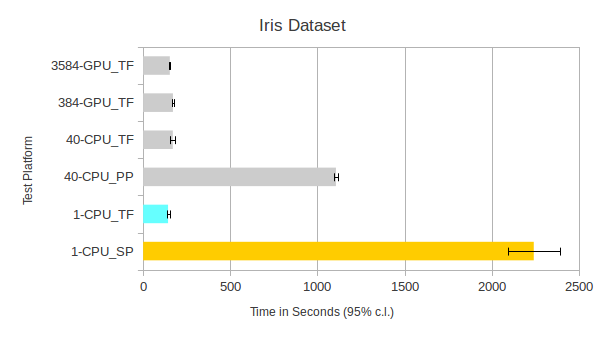}
\caption{A comparison of the average of 10 wall times (seconds) for the processing of the Iris dataset (150 x 4) against six test platform configurations (95\% cl). As with the Kepler dataset (Figure 1), the single CPU core comparison demonstrates the greatest improvement and overall fastest score. 1-CPU\_SP vs. 1-CPU\_TF results in a 15x improvement from scalar data and the SymPy maths library to vector data and the TensorFlow numerical computation library.}
\end{figure*}

\begin{figure*}[!h]
\centering \includegraphics[scale=0.8]{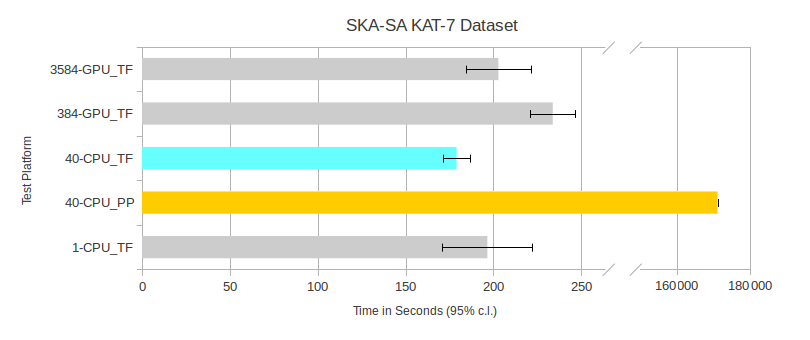}
\caption{A comparison of the average of 10 wall times (seconds) for the processing of the KAT-7 dataset against five test platform configurations (95\% cl). The original 12 runs were conducted in October 2015 with an average wall time of 48 hours (note the break in the x axis). In moving from processing scalar data through the SymPy maths and pprocess multicore libraries to vectorized data and the TensorFlow library, we see an 875x improvement in performance as demonstrated by 40-CPU\_PP vs. 40-CPU\_TF (same hardware). Overall, the SKA-SA 40 core Xeon server provides the highest performance on this (10,000 x 9) dataset.}
\end{figure*}

\begin{figure*}[!h]
\centering \includegraphics[scale=0.9]{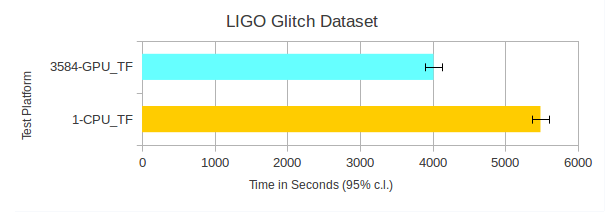}
\caption{A comparison of the average of 10 wall times (seconds) for the processing of the LIGO Glitch dataset against two test platform configurations (95\% cl). The fastest, overall test configuration is the LIGO GPU cluster node (3584-GPU\_TF) running TensorFlow. This demonstrates how larger datasets (millions of data points) benefit from the computational potential of a massively parallel engine.}
\end{figure*}

\begin{figure*}[!h]
\centering \includegraphics[scale=0.6]{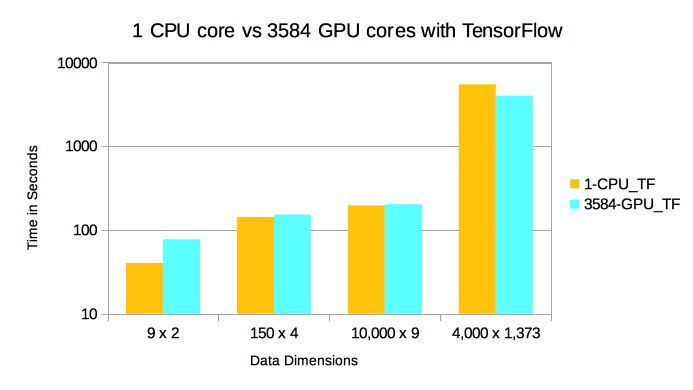}
\caption{A comparison of the average of 10 wall times (seconds) of two test platforms: 1-CPU\_TF and 3584-GPU\_TF against all four datasets. The benefit of the massively parallel GPU card is not seen until the platforms are processing the LIGO Glitch dataset with 5.5M total data points (4000 instances x 1373 features).}
\end{figure*}

\section{Acknowledgement}
We thank Professor Lee Spector, Hampshire College for his support; the Science and Technology Facilities Council, Newton Fund, and SKA South Africa for use of the KAT-7 data and server; the LIGO Scientific Collaboration for use of the glitch classification data; and the anonymous reviewers who provided valuable feedback to the improvement of this paper.  \newline

\bibliography{main}
\bibliographystyle{ACM-Reference-Format}

\end{document}